\documentclass[10pt,conference]{IEEEtran}
\usepackage{amsmath,amssymb,amsfonts}
\usepackage{hyperref}
\usepackage{algorithmic}
\usepackage{graphicx}
\usepackage{textcomp}
\usepackage{xcolor}
\usepackage{capt-of}
\usepackage{cuted}
\usepackage{booktabs}
\usepackage[numbers]{natbib}
\def\BibTeX{{\rm B\kern-.05em{\sc i\kern-.025em b}\kern-.08em
    T\kern-.1667em\lower.7ex\hbox{E}\kern-.125emX}}
\begin{document}

\title{TypeScript's Evolution: An Analysis of Feature Adoption Over Time
\thanks{This research was supported by an Australian Government Research Training Program Scholarship.}
}


\author{\IEEEauthorblockN{Joshua D. Scarsbrook, Mark Utting, Ryan K. L. Ko}
\IEEEauthorblockA{School of Information Technology and Electrical Engineering \\
The University of Queensland\\
Brisbane, Australia \\
\{j.scarsbrook,m.utting,ryan.ko\}@uq.edu.au}}

\maketitle

\begin{abstract}
TypeScript is a quickly evolving superset of JavaScript with active development of new features. Our paper seeks to understand how quickly these features are adopted by the developer community. Existing work in JavaScript shows the adoption of dynamic language features can be a major hindrance to static analysis. As TypeScript evolves the addition of features makes the underlying standard more and more difficult to keep up with. In our work we present an analysis of 454 open source TypeScript repositories and study the adoption of 13 language features over the past three years. We show that while new versions of the TypeScript compiler are aggressively adopted by the community, the same cannot be said for language features. While some experience strong growth others are rarely adopted by projects. Our work serves as a starting point for future study of the adoption of features in TypeScript. We also release our analysis and data gathering software as open source in the hope it helps the programming languages community.
\end{abstract}


\section{Introduction}

TypeScript\cite{TypeScript} is a fast evolving superset of JavaScript implementing static type checking. From 2020 to 2022, there have been ten releases each bringing additional features and most adding new syntax to the language. With the rapid pace of evolution the question becomes, how quickly are these features being picked up by the developer community? Are some features more popular than others?

This question has already been asked about other programming languages such as JavaScript, Java, and Python. In JavaScript the work by \citeauthor{10.1145/1806596.1806598}~\cite{10.1145/1806596.1806598} explores the use of dynamic language features in JavaScript and concludes that production applications often use dynamic features making static analysis challenging. Similar work has also been done in Java where \citeauthor{10.1145/1985441.1985446}~\cite{10.1145/1985441.1985446} discovered that most uses of generics were covered by a small number of classes but the usage varies between developers.

TypeScript has an evolving standard without a formal specification. Our paper seeks to understand how quickly new features in TypeScript are adopted, to determine how important it is for tools to stay up to date with the latest release. We hypothesize that it is unnecessary for program analysis tools to support the entire language and a smaller subset is sufficient for most applications.

In this paper, we focus specifically on syntactic features (features implemented in the Abstract Syntax Tree without modifying language semantics) introduced by TypeScript versions between 4.0 and 4.9 in popular TypeScript libraries and applications. TypeScript also sees regular improvements to type inference and language features that are expressed through the type checker. These features are not a focus for our study.

In this paper, we aimed to answer three research questions about the adoption of TypeScript features:

\begin{itemize}
    \item \textbf{(RQ1)} What are the most popular features recently introduced in TypeScript?
    \item \textbf{(RQ2)} How quickly are new TypeScript features adopted by projects that use TypeScript?
    \item \textbf{(RQ3)} How quickly are new TypeScript language versions adopted by projects that use TypeScript?
\end{itemize}

Results are presented in Section III. In this paper, we contribute:

\begin{itemize}
    \item A dataset of current popular TypeScript repositories collected from GitHub\cite{GitHub}.
    \item A open source framework for TypeScript feature/version adoption studies.
    \item The first study of the rate of language feature/version adoption for TypeScript.
    \item Recommendations for how important it is for tools to adopt new language features in TypeScript.
\end{itemize}

Our paper is organized into four further sections. We start with our methodology for analysis (Section II) before presenting our results (Section III). We then make a brief review of related work (Section IV) before concluding with a discussion including some future research directions (Section V).

\section{Methodology}

We ran our study on top-starred/rated repositories containing TypeScript code on GitHub. We extracted all commits between 2020 and 2022 (inclusive) and extracted a series of boolean flags indicating the usage of each language feature.  The analysis code and data sets used for this analysis are available in our repository.\footnote{See \url{https://github.com/Vbitz/jsdata\_msr}.}

\subsection{Dataset}

We started by downloading a list of the top $500$ TypeScript repositories from GitHub. The repositories are sorted according to the number of users that have starred the repository. These repositories include languages besides TypeScript code but only TypeScript is considered in our paper. We collected the list of repositories on January 4 2023 and included the list as part of our dataset. Our analysis includes all commits attached to a given repository. These projects often use feature branches which may include a feature well before its released on the main branch. For all calculations we used the date the feature first turned up in the repository rather than the date it was included in a release.

Of those $500$ repositories $23$ had no commits extracted and an additional $23$ recorded no versions of TypeScript. Therefore there are $454$ repositories with at least one version of TypeScript recorded.

\subsection{Analysis}

Our pipeline consists of an open source program written in Go\cite{Go} that extracts every unique TypeScript file from every commit in each repository. This includes all branches and all tags. We only consider commits made between January 1st 2020 and December 31st 2022 inclusive. We filtered by dates and only selected TypeScript features released between 2020 and 2022 because including commits outside this time span will not yield useful results.

In our dataset of $454$ repositories, $87\%$ contain less than $1000$ TypeScript files. We consider $1,325,810$ total commits in our analysis. Git commits contain multiple dates such as when the commit was authored versus when the commit was committed.  For our analysis we choose the latest possible date included in the commit.

Extracted TypeScript files are parsed by TypeScript and usage of different language features are detected according to their presence in the Abstract Syntax Tree (AST). With extensive caching and duplicate detection, the entire analysis takes approximately one hour.

\subsection{Version Detection}

We parse the \texttt{package.json} file in the root of the repository to detect the TypeScript version from the installed dependencies. 

\subsection{Feature List}

We focused on syntactic features which are exposed in the AST exported by TypeScript. We chose to focus on features released in the last three years (between 2020 and 2022).

TypeScript versions are released as Beta and a Release Candidate before they are formally released. In our paper, we consider the full release to be Day Zero, as listed below.  Projects adopting betas will show up as adopting features or versions before they were formally released
(a negative number of days relative to Day Zero). TypeScript 4.8 and 4.6 did not make syntactic changes to the language and only included semantic and inference changes.

\begin{table}
\centering
\caption{A list of the 8 TypeScript versions and 13 TypeScript features studied in our paper.}
\label{tab:typescript_features}
\begin{tabular}{@{}llll@{}}
\toprule
\textbf{Version} & \textbf{Release Date} & \textbf{} & \textbf{Name}                                     \\ \midrule
4.9              & 2022-11-15            & $f_{0}$   & \texttt{satisfies} operator                       \\
                 &                       & $f_{1}$   & \texttt{accessor} property                        \\ \midrule
4.7              & 2022-05-24            & $f_{2}$   & \texttt{extends} constraint on \texttt{infer}     \\
                 &                       & $f_{3}$   & Variance Annotations \texttt{in} and \texttt{out} \\ \midrule
4.5              & 2021-11-17            & $f_{4}$   & \texttt{type} import modifier                     \\
                 &                       & $f_{5}$   & Import assertions                                 \\ \midrule
4.4              & 2021-08-26            & $f_{6}$   & \texttt{static} blocks in classes                 \\ \midrule
4.3              & 2021-05-26            & $f_{7}$   & \texttt{override} modifier on methods             \\ \midrule
4.2              & 2021-02-23            & $f_{8}$   & Abstract constructs signature                     \\ \midrule
4.1              & 2020-11-19            & $f_{9}$   & Template literal types                            \\
                 &                       & $f_{10}$  & Key Remapping in Mapped Types                     \\ \midrule
4.0              & 2020-08-20            & $f_{11}$  & Labeled Tuple Elements                           \\
                 &                       & $f_{12}$  & Short-Circuiting Assignment                       \\ \bottomrule
\end{tabular}
\end{table}

Table \ref{tab:typescript_features} lists the $8$ versions and $13$ features in our study. This is not an exhaustive list of features introduced since we excluded features requiring type inference or type checking to identify.

Our dataset includes a few special repositories that have different characteristics to other projects:

\begin{itemize}
    \item TypeScript\cite{TypeScript}: The source code of TypeScript is included as part of this analysis.
    \item Babel\cite{Babel}: Babel is a compiler for JavaScript. It includes both ECMAScript\cite{ECMAScript} features and some TypeScript features since it has support for TypeScript syntax.
\end{itemize}

\section{Results}

To address our research questions, we started with the adoption of TypeScript versions before moving onto the adoption of TypeScript features.

\subsection{RQ1: Feature Adoption Rating}

We can categorize the adoption of features into two major groups based on their adoption slopes.

Group one contains $f_4$, $f_9$, $f_{11}$, $f_7$, $f_{12}$, $f_{10}$ and includes features that have more than $20$ repositories adopting them within one year after release. The most popular feature in this dataset is $f_4$ (\texttt{type} modifiers on \texttt{import}) and the second most popular is $f_9$ (Template Literal Types).

Both of these features were necessary to solve some missing gaps in the TypeScript language. \texttt{type} modifiers ensure imports are used only for type definitions, and are erased when the code is compiled. This allows including other libraries and files without including runtime dependencies. It can also help to break import loops in some cases where a module needs a type from a module that imports it. Template Literal Types similarly give more flexibility in how types are described and open up new avenues of meta programming.

Group two contains $f_2$, $f_1$, $f_8$, $f_3$, $f_0$, $f_5$, $f_6$ and includes features that have less than $20$ repositories implementing them one year after release. $f_6$ (static blocks in classes) has the lowest adoption rate among our dataset with only four repositories adopting it in a year.

Unlike $f_4$ and $f_9$ static blocks have equivalents in existing code so they are only used in niche circumstances.

\subsection{RQ2: Feature Adoption}

\begin{figure*}
    \centering
    \includegraphics[width=\linewidth]{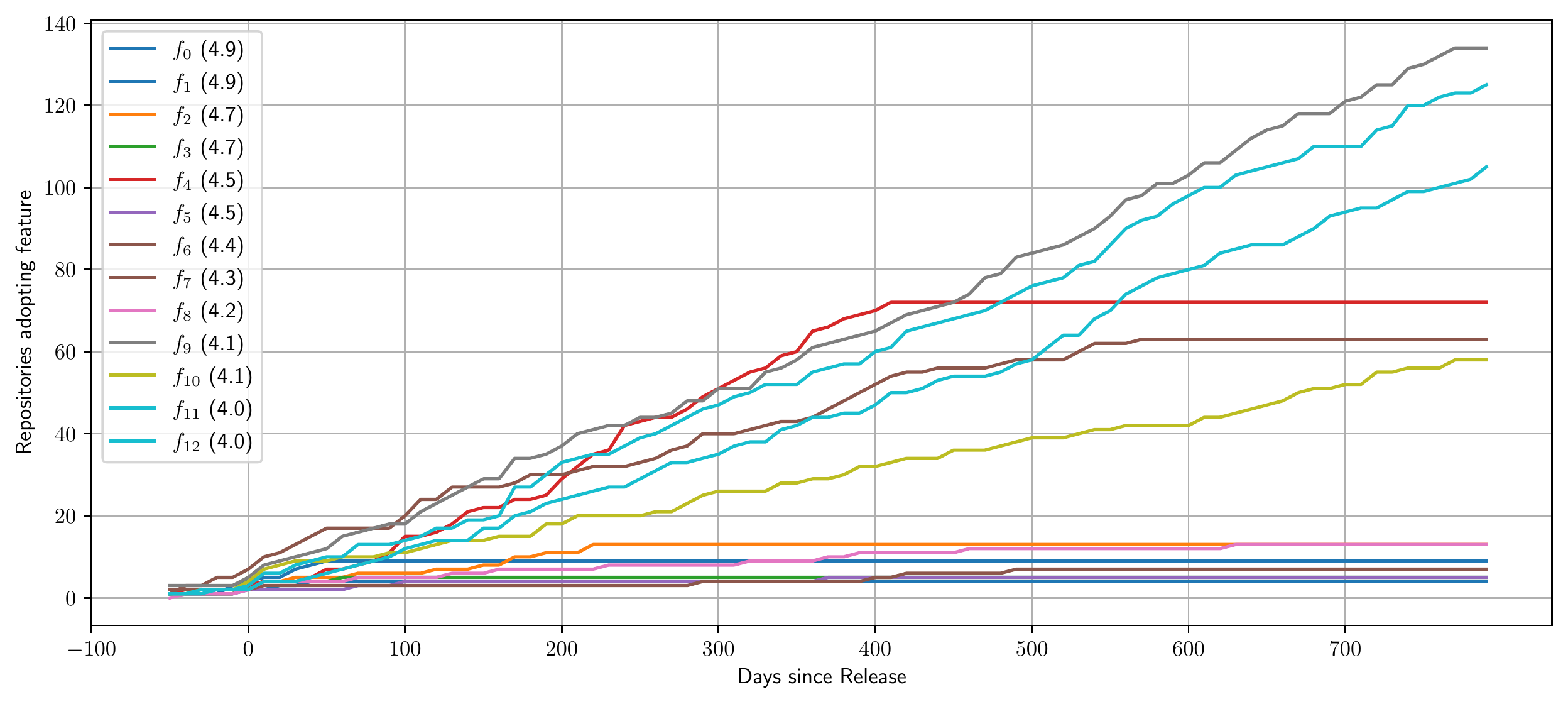}
    \caption{How quickly is each TypeScript feature adopted relative to one another. Note the release date of each feature as some features have not been released for all $800$ days.}
    \label{fig:feature_adoption}
\end{figure*}

Figure \ref{fig:feature_adoption} shows the adoption curve of each of the TypeScript feature we looked at. Unlike Figure \ref{fig:version_adoption} we can immediately see two major differences. Different features have significantly different adoption rates with some reaching high levels of adoption and some barely being adopted at all. Secondly all features have mostly linear adoption rates.

Features were detected across any file ending with \texttt{.ts} that can successfully be parsed as TypeScript. This means features that are only used in unit tests are also included here. In addition we include every branch of the repository so some features are adopted first in a feature branch before being included in the main branch.

\begin{table*}
    \centering
    \caption{The number of days before/after release where features were introduced into TypeScript and Babel.}
    \label{tab:ts_bab_features}
    \begin{tabular}{@{}rrrrrrrrrrrrrr@{}}
    \toprule
                        & $f_{0}$ & $f_{1}$ & $f_{2}$ & $f_{3}$ & $f_{4}$ & $f_{5}$ & $f_{6}$ & $f_{7}$ & $f_{8}$ & $f_{9}$ & $f_{10}$ & $f_{11}$ & $f_{12}$ \\ \midrule
\textbf{TypeScript} & $-73$     & $-80$     & $-80$     & $-62$     & $-50$    & $-57$     & $-61$    & $-60$     & $-46$     & $-70 $    & $-70$      & $-92$      & $-96$      \\
\textbf{Babel}      & $37$      & $-19$     & $-6$     & $-6$     & $-515$    &  N/A   & $-114$    & $-27$     & $-1$     & $-322$    & $-35$      & $-21$      & $211$      \\ \bottomrule
    \end{tabular}
\end{table*}

Both Babel and TypeScript were major outliers in the feature adoption rates. Table \ref{tab:ts_bab_features} focuses on these two repositories. Babel adopted some features well before TypeScript introduced them and TypeScript adopted all features before they were released. The behavior of TypeScript is easy to explain. A high coverage rate for unit tests means TypeScript starts adopting features as soon as they are implemented into the repository. Some features are part of the ECMAScript standard rather than TypeScript so Babel may include these features before TypeScript adds support for them. That explains why Babel adopts some features well before TypeScript.

\subsection{RQ3: TypeScript Versions}

\begin{figure}
    \centering
    \includegraphics[width=\linewidth]{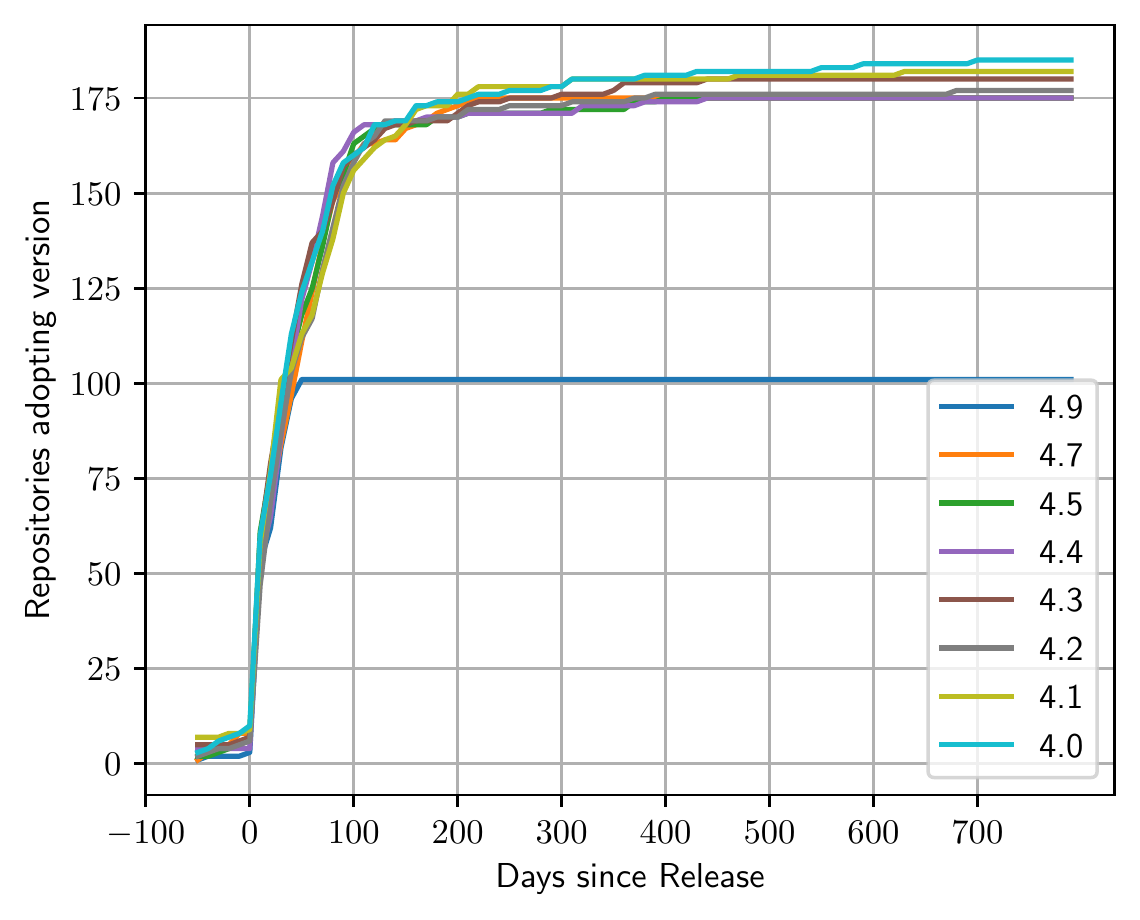}
    \caption{The adoption curves of different versions. Version 4.9 was released 50 days before the data collection ended (31st December 2022) so the data stops there.  Adoption rates asympote to 180 projects, which is around $40\%$ of projects.  Other projects jump versions, rather than adopting every version.} 
    \label{fig:version_adoption}
\end{figure}

Figure \ref{fig:version_adoption} shows the adoption curve of each of the TypeScript versions we pulled features from. We can see here that all versions follow a similar adoption curve, with an initial slow adoption of pre-release versions, then a rapid adoption in the three months after release, followed by slower late adoption by a small number of repositories.  Roughly $1/3$ of projects ($160$ out of $454$) adopt the latest release within the first three months after release (except for TypeScript 4.9, which was released less than 50 days before data collection ended). These fast adoption curves are not surprising since JavaScript/TypeScript projects regularly update dependencies to the latest revision and TypeScript releases do not introduce significant breaking changes.

Most adoption happens in the first three months after release (Roughly $35\%$ of projects in our dataset) with a small tail at the end for projects that update after a new version is already released. At the time of writing, TypeScript releases new versions every three months so some projects may not have adopted a version before the new version is released.

These are results aggregated over $454$ different repositories and we do not see all repositories accounted for here. This comes down to two major reasons. While some repositories (Visual Studio Code for example) adopt new versions within a few  days of release some take a few months to adopt new versions or do not adopt them at all. The adoption averages out to the same curve though. The other reason is we detected the TypeScript version using the \texttt{package.json} file. The maximum adoption for any version is $185$ out of $454$ ($46$ have no TypeScript version recorded). The reason is not all repositories adopt all versions of TypeScript and most skip versions as they do not regularly update.

A few repositories adopted versions before they were formally released. TypeScript depends on itself but overrides that with the local version. It therefore adopts new versions before they are released. 

\section{Related Work}

Some existing work has already investigated adoption of language features in JavaScript \cite{10.1145/1806596.1806598,10.1002/spe.2334,10.1007/978-3-642-22655-7_4}, Java \cite{10.1145/1985441.1985446,10.1145/2568225.2568295,10.1007/978-3-540-70592-5_28}, and Python \cite{9425916,10.1145/2597073.2597103}.

JavaScript allows for self-modifying code and code generated and evaluated at runtime. These features make tracking the control flow over a programs execution difficult so some previous works exclude them from analysis. The work by \citeauthor{10.1145/1806596.1806598}~\cite{10.1145/1806596.1806598} questions this approach by looking at the prevalence of these features in production code. Due to the nature of those features most analysis there is based on dynamic analysis rather than the static analysis we use in our work. 

A large amount of work in this area has been done in Java \cite{10.1145/1985441.1985446,10.1145/2568225.2568295,10.1007/978-3-540-70592-5_28}. Firstly the work by \citeauthor{10.1145/1985441.1985446}~\cite{10.1145/1985441.1985446} discovered that most uses of generics were covered by a small number of classes but the usage varies between developers. The work by \citeauthor{10.1145/2568225.2568295}~\cite{10.1145/2568225.2568295} broadened this by looking at $31,432$ Java projects on SourceForge \cite{SourceForge} and studying the adoption of $18$ language features introduced in three versions of Java.

The work by \citeauthor{9425916}~\cite{9425916} performs a similar study to our work focusing on Python projects instead of TypeScript projects. They perform a smaller study on $35$ different projects across a range of sectors. They make the interesting observation that larger projects tend to use less involved language features like safety checks rather than more advanced features like diamond inheritance. This lines up with our outcome since the most popular features we observed increase safety and the least popular feature (static blocks in classes) can make control flow more difficult to read. Another work by \citeauthor{10.1145/3524842.3528467}~\cite{10.1145/3524842.3528467} follows a similar direction to the work by \citeauthor{10.1145/1806596.1806598}~\cite{10.1145/1806596.1806598} looking at the impact of dynamic features on static analysis of Python code.

The work by \citeauthor{10.1145/3475738.3480941}\cite{10.1145/3475738.3480941} includes a brief analysis of feature usage in DefinitelyTyped\cite{DefinitelyTyped}. The work is limited to types in type declaration files rather than our study looking at TypeScript source code.

The static analysis field leverages this study to inform the language features they implement support for. For instance the work by \citeauthor{10.1145/2676726.2676971}\cite{10.1145/2676726.2676971} seeks to improve the safety of TypeScript programs and uses a smaller subset of TypeScript called "Safe TypeScript". This work was done before prior to the release of TypeScript 1.1 (October 6, 2014) and lacks may of the features introduced after. In addition the work by \citeauthor{10.1145/2714064.2660215}\cite{10.1145/2714064.2660215} uses a version of the TypeScript language to detect faults in JavaScript interfaces. Like the work by \citeauthor{10.1145/3475738.3480941}\cite{10.1145/3475738.3480941} it focuses on declaration files rather than TypeScript source code.

Overall the related work covers two different kinds of study. Some work \cite{10.1145/1806596.1806598} uses dynamic analysis to study the prevalence of dynamic features. The other group of studies \cite{9425916} look at the usage of features across different types of project. Another further field\cite{10.1145/3475738.3480941,10.1145/2714064.2660215,10.1145/2676726.2676971} uses static analysis to perform code analysis on TypeScript language features. Our work extends on the second field of work by looking at a series of different versions.

\section{Discussion \& Concluding Remarks}

The answer to \textbf{RQ1} is that the most popular new language features are \texttt{type} modifiers on \texttt{import}s and template literal types. While \texttt{type} modifiers solve an existing issue of unintended side effects from imported modules, template literal types give additional flexibility in how types are constructed.

The answer to \textbf{RQ2} is more involved. Different features are adopted at different rates which is an expected outcome. Some features are very niche and are only used by a small number of libraries. The unexpected outcome is that adoption rates are static over time and no features sees a large initial peak as developers race to adopt them. Our interpretation of this is that very few projects need a new feature, so they are adopted as developers learn about them and gradually utilize them in new code and in code rewrites.

Finally, the answer to \textbf{RQ3} is straightforward. Most projects adopt new versions of TypeScript quickly with an expected long tail as remaining projects update to new versions.

\subsection{Conclusions}

We observed a simple adoption curve for language versions, with most adoption happening shortly after release with ${1}/{3}$ of repositories updating before the next TypeScript version is released. However, the adoption of new language features into repositories is much more gradual. A project can update to a new version of TypeScript without changing their code at all, so without adopting any new features.  So adopting a new language feature may require adopting a new TypeScript version, but not vice versa.  We can draw the conclusion that while a project has a feature available it may not adopt it until much later.

Returning to our overall goal of specifying a useful subset of TypeScript for program analysis tools we can see that although new language versions are adopted quickly by the ecosystem (${1}/{3}$ over $3$ months) the adoption of new features is a lot more variable with some features never being adopted outside of a few projects. This shows that it is important for tools to keep up to date with language versions but it is less important to support all language features (e.g. Group 2 features are used by only a few projects).

\subsection{Future Work}

Currently our analysis focuses on syntactic changes to TypeScript, which misses improvements made to type inference and to the developer experience. It would be useful in future research to expand the list of features and look at semantic changes.

Our paper focuses on the features introduced in the 4.x versions of TypeScript to make timely analysis possible. Future work could look at additional TypeScript versions.

Additionally it would be interesting to run our analysis on a wider body of repositories to see how the results change with less popular projects.

\bibliography{references}

\end{document}